\documentclass[nofootinbib,preprint,aps,showpacs]{revtex4}
\usepackage{graphicx}
\usepackage{dcolumn}
\usepackage{bm}

\begin{document}

\preprint{}

\title{Scale-dependence of transverse momentum correlations in Pb-Au collisions at 
158$A$ GeV/$c$}

\def\rez{$^{1}$}
\def\gsi{$^{2}$}
\def\fra{$^{3}$}
\def\dub{$^{4}$}
\def\hei{$^{5}$}
\def\wei{$^{6}$}
\def\sun{$^{7}$}
\def\cer{$^{8}$}
\def\mpi{$^{9}$}
\def\bnl{$^{10}$}
\def\mun{$^{11}$}

\author{
D.~Adamov\'a\rez, 
G.~Agakichiev\gsi, 
D.~Anto\'nczyk\gsi,
H.~Appelsh\"auser\fra, 
V.~Belaga\dub, 
S.~Bielcikova\hei,
P.~Braun-Munzinger\gsi,
O.~Busch\gsi,
A.~Cherlin\wei, 
S.~Damjanovi\'c\hei, 
T.~Dietel\hei, 
L.~Dietrich\hei, 
A.~Drees\sun,
W.~Dubitzky\hei,
 S.\,I.~Esumi\hei, 
K.~Filimonov\hei, 
K.~Fomenko\dub,
Z.~Fraenkel\wei$^{,\dagger}$, \footnotetext{$^{\dagger}$deceased} 
C.~Garabatos\gsi, 
P.~Gl\"assel\hei, 
J.~Holeczek\gsi,
V.~Kushpil\rez, 
A.~Maas\gsi, 
A.~Mar\'{\i}n\gsi, 
J.~Milo\v{s}evi\'c\hei,
A.~Milov\wei, 
D.~Mi\'skowiec\gsi, 
Yu.~Panebrattsev\dub, 
O.~Petchenova\dub, 
V.~Petr\'a\v{c}ek\hei, 
A.~Pfeiffer\cer,
M.~P\l{}osko\'n\fra, 
S.~Radomski\gsi,
J.~Rak\mpi, 
I.~Ravinovich\wei, 
P.~Rehak\bnl,
H.~Sako\gsi, 
W.~Schmitz\hei, 
S.~Sedykh\gsi, 
S.~Shimansky\dub, 
J.~Stachel\hei, 
M.~\v{S}umbera\rez, 
H.~Tilsner\hei, 
I.~Tserruya\wei, 
G.~Tsiledakis\gsi,
J.\,P.~Wessels\mun, 
T.~Wienold\hei, 
J.\,P.~Wurm\mpi, 
W.~Xie\wei, 
S.~Yurevich\hei, 
V.~Yurevich\dub \\
~\\
(CERES Collaboration)\\
~\\
}

\address{
\rez Nuclear Physics Institute ASCR, 25068 \v{R}e\v{z}, Czech Republic\\
\gsi Gesellschaft~f\"{u}r~Schwerionenforschung~(GSI),~D-64291~Darmstadt,~Germany\\
\fra Institut f\"{u}r Kernphysik der Universit\"{a}t Frankfurt,~D-60438 Frankfurt,~Germany\\
\dub Joint Institute for Nuclear Research, 141980 Dubna, Russia\\
\hei Physikalisches Institut der Universit\"{a}t Heidelberg, D-69120 
Heidelberg, Germany\\
\wei Weizmann Institute, Rehovot 76100, Israel\\
\sun Department of Physics and Astronomy, State University of 
New York--Stony Brook, Stony Brook, New York 11794-3800\\
\cer CERN, 1211 Geneva 23, Switzerland\\
\mpi Max-Planck-Institut f\"{u}r Kernphysik, D-69117 Heidelberg, Germany\\
\bnl Brookhaven National Laboratory, Upton, New York 11973-5000\\
\mun Institut f\"{u}r Kernphysik der Universit\"{a}t M\"unster, D-48149 M\"unster,~Germany\\
}


\begin{abstract}
We present results on transverse momentum correlations of charged particle
pairs produced in Pb-Au collisions at 158$A$ GeV/$c$ at the Super Proton Synchrotron.
The transverse momentum correlations have been studied as a function of collision
centrality, angular separation of the particle pairs, transverse momentum 
and charge sign. 
We demonstrate that the results are in agreement with previous findings in
scale-independent analyses at the same beam energy. Employing the two-particle
momentum correlator $\langle \Delta p_{t,i}, \Delta p_{t,j}\rangle$ and the 
cumulative $p_t$ variable $x(p_t)$, we identify, using the scale-dependent 
approach presented in this paper, different sources contributing to the measured
correlations, such as quantum and Coulomb correlations, elliptic
flow and mini-jet fragmentation. 
\end{abstract}

\maketitle

\def\corr{\langle \Delta p_{t,i}, \Delta p_{t,j}\rangle}
\def\corrdiff{\corr (\Delta \eta, \Delta \phi)}

\section{\label{sec:intro} Introduction}

Strongly interacting matter is expected to exist in different phases.
At high temperature and vanishing density, QCD calculations on the lattice 
indicate the formation of a deconfined system with partonic 
degrees of freedom, the Quark-Gluon Plasma (QGP)~(see \cite{lat} for a 
recent review).
The QGP is separated from the hadronic phase by a transition
line of yet unknown order. 
Lattice QCD calculations at finite baryon chemical potential $\mu_B$ are 
more difficult, but substantial progress has been achieved 
recently~\cite{lat}. 
While the transition is most likely a rapid cross-over at vanishing density,
it is expected to be first order at large $\mu_B$. The conjecture of the 
existence a 
critical point arises naturally in this configuration, however, non-standard
scenarios without critical endpoint have also been discussed~\cite{philipsen}.

Collisions of heavy nuclei at high energies provide experimental
access to the phase structure of hot and dense nuclear matter.
If thermalization is achieved in the early stage of the collision,
a phase transition to a Quark-Gluon-Plasma may occur. 
Experimental signatures of the phase transition and QGP formation 
are often based on event-averaged distributions of final state particles. 
Additional information may be derived from the study of
fluctuations, which are related to the thermodynamic
properties of the system (for a review see \cite{jeon+koch}).
Event-by-event fluctuations of
intensive quantities like temperature, reflected in the 
mean transverse momentum $M_{pt}$
of hadrons, have been proposed as a possible signature for 
critical phenomena connected with the passage of the system
through the phase boundary, and may therefore constitute an independent 
probe to pin down the properties of the QCD phase 
diagram~\cite{shur,steph1,steph2,heisel,dum,mro,fodor}.

Fluctuations of the mean transverse momentum are associated to the
heat capacity of the system, which has a maximum at the QCD phase
boundary. As a conseqence, transverse momentum fluctuations in the 
final state could be suppressed if the system retains its memory from
the phase transition. 
On the other hand, long-range correlations associated with the vicinity
of the system to the QCD critical point may lead to an enhancement of
transverse momentum fluctuations.

Recent measurements of event-by-event fluctuations of the mean transverse
momentum at SPS and RHIC demonstrated a clear excess beyond the statistical
expectation~\cite{ceres-1,ceres-2,na49-1,star-1,star-2,phenix-1}. 
Different regions of the QCD phase diagram have been probed by variation
of experimental control parameters, such as beam energy and system size.
It was argued that passage of the system through the critical point or 
the phase boundary would be signaled by a sudden change of the fluctuation
pattern. While the energy dependence turned out to be structureless within
the SPS and RHIC energy regime,
a characteristic centrality and system size
dependence has been observed, exhibiting a non-monotonic behaviour with
a maximum at system sizes corresponding to about 100-150 participating
nucleons.
However, a strong connection of these findings to
critical behaviour in the vicinity of the phase transition 
has not yet been made.

This inconclusive result is partially caused by the 
difficulty to identify and disentangle various
contributions to the
different fluctuation measures. While the implications of finite
number statistics are well under control, there are a number of
sources of 
correlations on different transverse momentum scales,
which are expected to contribute to the fluctuation measures, such 
as quantum statistics,
final state interactions, collective flow, 
resonance decays or jet fragmentation.
Typically, the sensitivity to these contributions is tested with the
help of Monte-Carlo studies.

In this paper, we follow a novel approach by studying the scale-dependence
of transverse momentum correlations. The aim is to identify and separate 
different sources of correlations by a differential, scale-dependent analysis
of the two-particle correlation pattern. 
A similar approach to a scale-dependent correlation analysis has been
followed at RHIC, using different variables and an elaborated inversion
technique~\cite{tt1,tt2}.
The formalism used in this paper
allows to relate the measured differential correlation strength to the global
fluctuation measures used in previous analysis, such as 
$\Phi_{p_t}$~\cite{gaz-1},
$\sigma_{\rm dyn}^{2}$~\cite{volo-1}, or $\Sigma_{p_t}$~\cite{ceres-1}.
The present results are based on an analysis of a
high statistics data set of Pb-Au collisions at 
158$A$ GeV/$c$
recorded with the Time
Projection Chamber of the CERES experiment at CERN-SPS.

\section{\label{sec:meas} Measures of correlations and fluctuations}

Fluctuations of the event-by-event mean transverse momentum
$M_{pt}$ are composed of statistical fluctuations arising
from the finite number of tracks per event, and a possible non-statistical
({\em dynamical}) contribution.
The mean transverse momentum $M_{pt,k}$ of event $k$ with $N_k$
charged particles is defined as 
\begin{equation}
	M_{pt,k} = \frac{\sum_{i=1}^{N_k}p_{t,i}}{N_k}.
\end{equation}
As a measure for event-by-event fluctuations, we 
employ the dynamical fluctuation $\sigma^2_{p_t,\rm dyn}$~\cite{volo-1} 
that is derived from the
variance of the inclusive single track $p_t$ distribution
$\overline{\Delta p_t^2}$, the variance of the event-by-event
mean transverse momentum distribution $\langle \Delta M^2_{pt}\rangle$
and the average number $\langle N \rangle$ of charged particles   
per event:
\begin{equation}
	\sigma^2_{p_t,\rm dyn} = \langle \Delta M^2_{pt}\rangle -
	\frac{\overline{\Delta p_t^2}}{\langle N \rangle}.	
\end{equation}
For the calculation of the variance $\langle \Delta M^2_{pt}\rangle$,
the mean transverse momentum $M_{pt,k}$ of each event $k$ has been weighted
by the number of tracks $N_k$ in this event.
The dynamical fluctuation $\sigma^2_{p_t,\rm dyn}$ is zero
by definition if all fluctuations are purely statistical.
For convenience, the normalized dynamical fluctuation $\Sigma_{pt}$
is introduced~\cite{ceres-1}:
\begin{equation}
  \Sigma_{pt} = {\rm sgn}\left(\sigma^2_{p_t,\rm dyn}\right)
	\frac{\sqrt{|\sigma^2_{p_t,\rm dyn}|}}{\overline{p_t}},
\end{equation}
where $\overline{p_t}$ is the inclusive mean transverse momentum 
of all tracks from all events.
The measure $\Sigma_{pt}$ is dimensionless and specifies the 
dynamical contribution to event-by-event $M_{pt}$ fluctuations
in units of $\overline{p_t}$. For the case of independent
particle emission from a single parent distribution, $\Sigma_{pt}$ vanishes.

Most of the previous studies of $M_{pt}$ fluctuations have been performed
in a scale-independent way, integrating over all short- and long-range
contributions present in the detector acceptance.
For such a scale-independent approach, the measures presented above
are well suited. 
However, more details about the origin of non-statistical 
fluctuations can be obtained by the study of their scale dependence.
For this purpose, an additional measure will be introduced.

The occurence of non-statistical fluctuations of $M_{pt}$ goes along 
with correlations among the transverse momenta of particles. 
Such correlations can be calculated
employing the two-particle transverse momentum correlator~\cite{star-2}:
\begin{equation}
\corr=
\frac{1} 
{\sum_{k=1}^{n_{\rm ev}}N_k^{\rm pairs}}.
\sum_{k=1}^{n_{\rm ev}}
\sum_{i=1}^{N_k}\sum_{j=i+1}^{N_k}(p_{ti}-\overline{p_t})
(p_{tj}-\overline{p_t}) .
\end{equation}
As for the measures defined earlier, $\corr$ 
can be calculated for distinct classes of tracks after
application of single track cuts, like cuts on $p_t$,
pseudorapidity or charge sign.
In this case,
the number of pairs in event $k$ is given by 
$N_k^{\rm pairs} = 0.5 \cdot N_k(N_k-1)$, where 
$N_k$ is the number of particles of a given class in event $k$,
and $\corr$ is approximately equal to 
$\sigma^2_{p_t,\rm dyn}$.

Since $\corr$ is a particle pair observable, it 
is possible to apply cuts on the pair level, contrary to the situation 
for the measures introduced above. An example is
the study of $p_t$ correlations between particles of different
charge sign. In this case, the correlator is calculated in the
following way:
\begin{equation}
\corr^{(+-)} = 
\frac{1}
{\sum_{k=1}^{n_{\rm ev}}N_k^{+}N_k^-}.
\sum_{k=1}^{n_{\rm ev}}
\sum_{i=1}^{N_k^+}\sum_{j=1}^{N_k^-}(p_{ti}-\overline{p_t}^{(+)})
(p_{tj}-\overline{p_t}^{(-)}) .
\end{equation}
Moreover, it is also possible to study the scale dependence
of $p_t$ correlations in angular space by calculating the correlator in bins
of the angular separation $\Delta \eta, \Delta \phi$ of particle 
pairs: 
\begin{equation}
\corrdiff =
\frac{1}
{\sum_{k=1}^{n_{\rm ev}}N_k^{\rm pairs}(\Delta \eta, \Delta \phi)}.
\sum_{k=1}^{n_{\rm ev}}
\sum_{i=1}^{N_k}\sum_{j=i+1}^{N_k}(p_{ti}-\overline{p_t})
(p_{tj}-\overline{p_t}) .
\end{equation}
In this case, only particle pairs with a given $\Delta \eta, \Delta \phi$
contribute to the correlator sum. 

\section{\label{sec:exp} Experiment and Data Analysis}
The CERES experiment at the CERN Super Proton Synchrotron (SPS)
has been designed for the study of electron-pair production
at mid-rapidity in hadronic and nuclear collisions.
By the addition of a cylindrical Time Projection Chamber (TPC) 
and a new magnet system in 1998, the momentum resolution of the 
spectrometer was significantly improved~\cite{ana-qm04}. Moreover, the
excellent tracking capability of the TPC for charged particles
provides the opportunity to study also hadronic observables.
The TPC is located about 4~m downstream of a segmented gold target and
covers a polar angle range from $7^{\circ}$ 
to $15^{\circ}$ at full azimuth. 
Charged particles create a trace of ionization in the Ne-CO$_2$ 
(80:20) gas mixture. Ionization electrons drift radially outwards 
to the read-out plane where up to 20 subsequent hits are detected
along the particle track.
Taking the drift time into account, the TPC allows a three-dimensional
reconstruction of the charged particle trajectory.
The tracking efficiency in the TPC is about 85\% for transverse
momenta greater than 0.1~GeV/$c$. 
In azimuth, the read-out plane is subdivided into 16 read-out chambers,
arranged along the polygon-like outer perimeter of the TPC.
Efficiency losses occur predominantly for
tracks which cross between two adjacent read-out chambers.

The TPC is operated inside an inhomogenous 
magnetic field with a radial component
of up to 0.75~T.
By the measurement of its deflection,
the momentum of a charged particle can be reconstructed 
with a resolution of 
$\Delta p/p = 2\% \oplus 1\%\cdot p~$(GeV/$c$)~\cite{ana-qm04}. 

In the year 2000, a sample of $3.2\cdot10^7$ central 
($\sigma/\sigma_{\rm geo} \approx 8\%$) and about $10^6$ minimum 
bias Pb-Au events have been recorded. 
The centrality selection is based on the pulse height
in a scintillator Multiplicity Counter (MC)~\cite{darek-qm05}. 
By the use of a geometric 
nuclear overlap model~\cite{eskola}, the centrality can be expressed in
terms of the number $\langle N_{\rm part}\rangle$ of nucleons participating 
in the collision.

The analysis presented here is based on $10^7$ central and
$2\cdot10^6$ min bias Pb-Au events at 158$A$ GeV/$c$ from the 
year 2000~\cite{georgios}. 
We use charged particles reconstructed in the TPC in the kinematic
range $2.2<\eta<2.7$ and $0.1<p_t<1.5$~GeV/$c$. 
A minimum of 12 
detected hits per track is required
to provide sufficient momentum resolution. 
Background from decays and conversions is suppressed by
the requirement that the back-extrapolation of the TPC tracks
into the target plane should not miss the interaction point by more than
10~cm in transverse direction.

In the scale-independent analysis, the measures $\sigma^2_{p_t,\rm dyn}$,
$\Sigma_{p_t}$, and $\corr$ have been calculated for positive
and negative particles separately, as well as without charge selection.
In addition, transverse momentum correlations between particles with opposite 
charge sign are analyzed employing the measure $\corr^{(+-)}$.

For the study of the scale dependence, the momentum correlator
$\corrdiff$
has been evaluated in bins of the relative angular separation 
$\Delta \phi = |\phi_i - \phi_j|$ and $\Delta \eta = |\eta_i - \eta_j|$
of the particle pair. We have chosen bins of 7.5 degrees in azimuth and
0.1 in pseudorapidity which corresponds to approximately equal 
granularity in polar and azimuthal angle. 

\begin{figure}
\centering
\includegraphics[height=12cm]{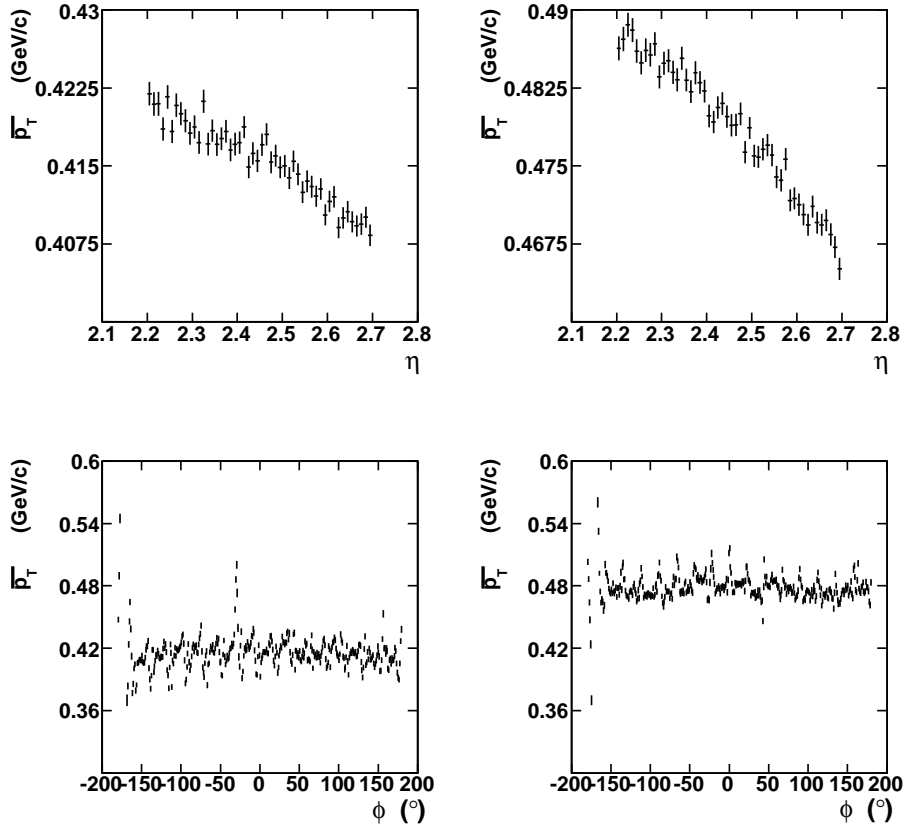}
\caption{Mean transverse momentum as function of $\eta$ and $\phi$.
Negative particles (left) and positive particles (right) are shown separately.}
\label{fig1}       
\end{figure}

Fig.\ref{fig1} shows the inclusive mean transverse momentum $\overline{p_t}$ 
measured in the TPC as function of $\eta$ and $\phi$. 
A decrease of $\overline{p_t}$ 
by a few percent 
as function of pseudorapidity is observed. 
The decrease can be explained by the kinematical acceptance
of the TPC, combined with a pseudorapidity-dependent 
proton-to-pion ratio. This has been verified by a Monte-Carlo
simulation.  
In azimuth, the 16-fold 
structure of the TPC read-out plane is visible. This effect arises
as a consequence of efficiency losses in the vicinity of the gaps
between adjacent read-out chambers. In detail, efficency losses
depend on the curvature of the track, hence giving rise to local
variations of $\overline{p_t}$.

If a correlation analysis is performed
in bins of relative angular space,
such variations lead to trivial (anti-) correlations of transverse
momenta 
as function 
of $\Delta \phi$ and $\Delta \eta$.
A correction of the final results 
for these correlations can be obtained by a mixed-event procedure.
Pairs of particles from different events are free of correlations
except those arising from the single particle acceptance of the
detector. 
Mixed-event correlations can therefore be used as a reference 
to correct the same-event results for such effects.

\begin{figure}
\centering
\includegraphics[height=12cm]{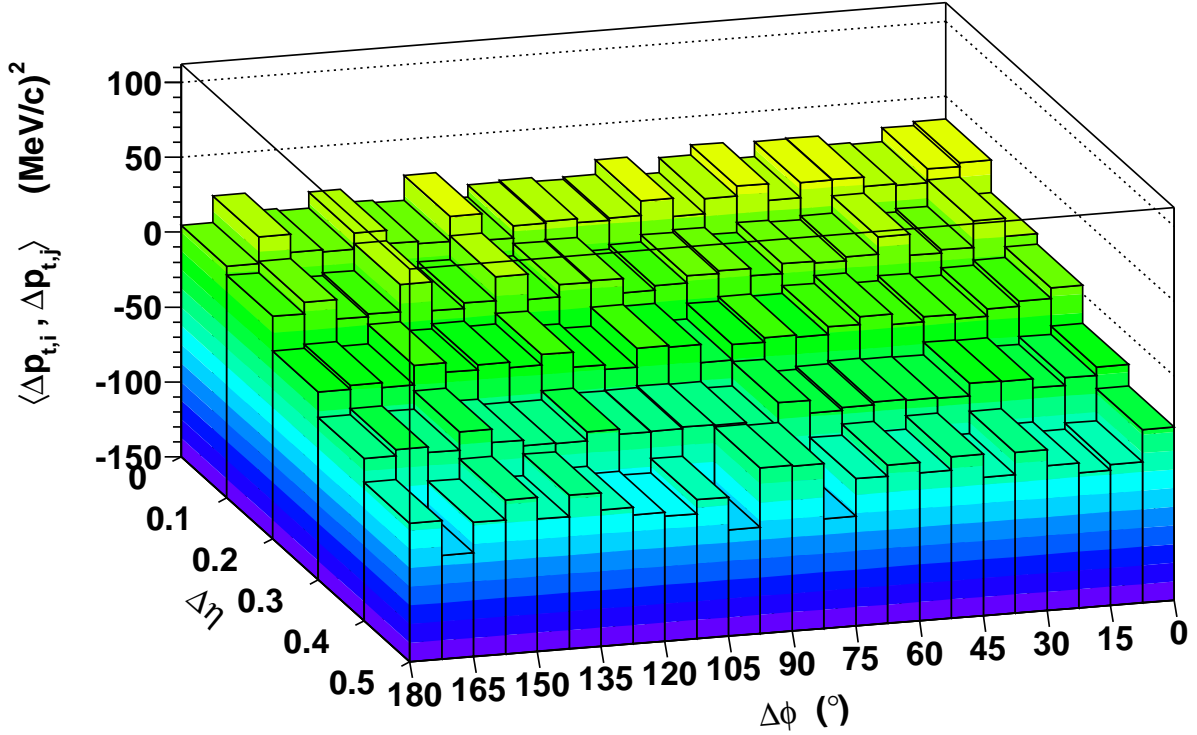}
\caption{Result of a momentum correlator analysis of mixed-event particle 
pairs.}
\label{fig2}       
\end{figure}

The result for the momentum correlator in the $\Delta \phi - \Delta \eta$
plane for mixed-event pairs is shown in Fig.\ref{fig2}. Statistical
errors range from about 2~MeV$^2$/c$^2$ in the first $\Delta \eta$ bin
to about 7~MeV$^2$/c$^2$ in the last $\Delta \eta$ bin. In $\Delta \phi$
direction, the observed variations are small. This is expected from
the high frequency pattern in Fig.\ref{fig1} (right panel): By averaging
over all combinations, correlations and anti-correlations 
cancel to a large extent. 
In Fig.\ref{fig2}, only a residual structure of the read-out 
plane segmentation is visible, showing up as a weak repetitive pattern 
in azimuthal direction. The wave length of the structure is 3 bins 
or $22.5^{\circ}$ in
Fig.\ref{fig2}, corresponding to $360^{\circ}/16$, the azimuthal size
of one TPC read-out chamber.

The situation is different in $\Delta \eta$. Due to the monotonic 
decrease of $\overline{p_t}$ as function of pseudorapidity, pairs 
are preferentially correlated at small $\Delta \eta$ and anti-correlated
at large $\Delta \eta$. This is demonstrated in Fig.\ref{fig2}, where 
a significant monotonic decrease of the correlator as function
of $\Delta \eta$ can be observed.
 
The momentum correlator has been calculated for mixed-event
pairs as function of $\Delta \eta$ and $\Delta \phi$ and for all
possible charge combinations. These results have been subtracted
bin-wise from the corresponding results of the same-event 
analyses presented below. 

It should be noted 
that, in scale-independent analyses of the momentum correlator, correlations
due to the single track $p_t$ acceptance cancel. The result of a
scale-independent mixed-event analysis averaging over all relative
angles yields $\corr_{\rm mixed} = 
-0.049\pm0.314$~MeV$^2$/$c^2$, although the differential analysis
(Fig.\ref{fig2}) shows significantly non-zero results.

Statistical errors are calculated by division of the total
event sample into $N_{\rm sub}$ subsamples of equal size. 
The variance $V$ of the results 
from the subsample analysis was used to determine the final
statistical errors for the full sample given by $\sqrt{V/N_{\rm sub}}$.
The systematic uncertainties have been estimated by variations
of the cut values and yield about 5~MeV$^2$/$c^2$ in $\corr$.

\section{Results}
\subsection{Scale-independent analysis}
In Table~\ref{tab:tab1}, results 
are shown of a scale-independent analysis
of about $10^7$ central Pb-Au events at 158$A$ GeV/$c$ 
with $\sigma/\sigma_{\rm geo}=0-8\%$.
The measures $\sigma^2_{p_t,\rm dyn}$, $\Sigma_{p_t}$
and $\corr$ have been calculated with and without
charge selection. Results for $\corr^{(+-)}$ using
pairs of different charge sign are also shown.

\begin{figure}
\centering
\includegraphics[height=12cm]{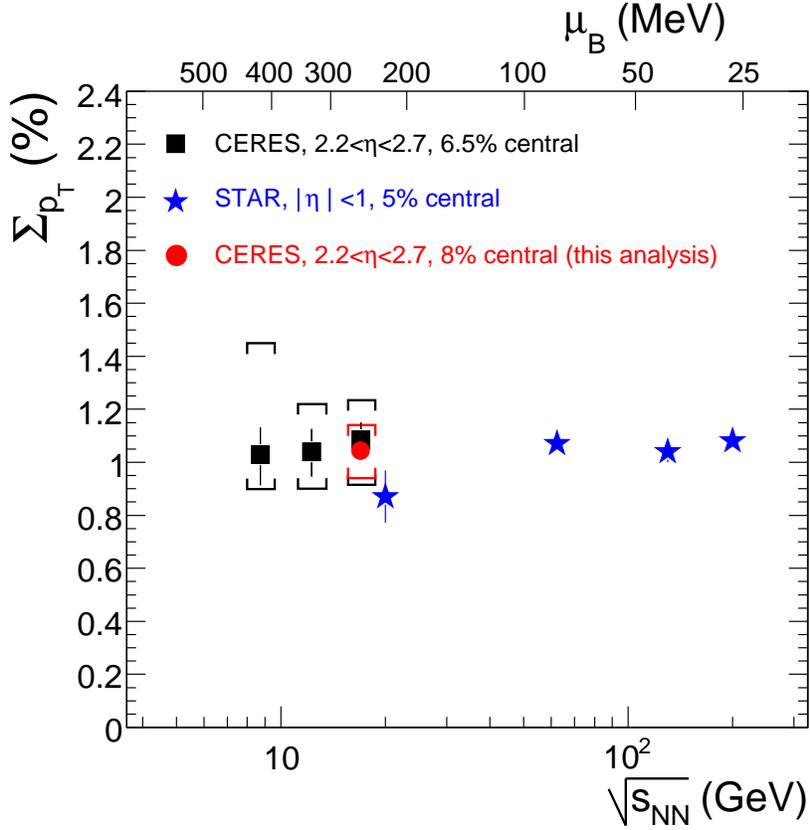}
\caption{Normalized dynamical fluctuation $\Sigma_{p_t}$ in central 
Pb-Au and Au-Au collisions as function of $\sqrt{s}$.}
\label{fig-roots}       
\end{figure}

For a given charge combination, the results obtained for 
$\corr$ and $\sigma^2_{p_t,\rm dyn}$ are consistent
within statistical errors.
For negative particles, the results are slightly higher than for positive
particles. Momentum correlations between unlike-sign particles
are comparable to those between particles of the same charge.
The normalized dynamical fluctuation $\Sigma_{p_t}$ is of order
1\% of mean $p_t$ in all cases, in good agreement with previously
reported measurements in central events at SPS and RHIC 
(see Fig.\ref{fig-roots})~\cite{ceres-1,ceres-2,star-2}.

\begin{figure}
\centering
\includegraphics[height=6cm]{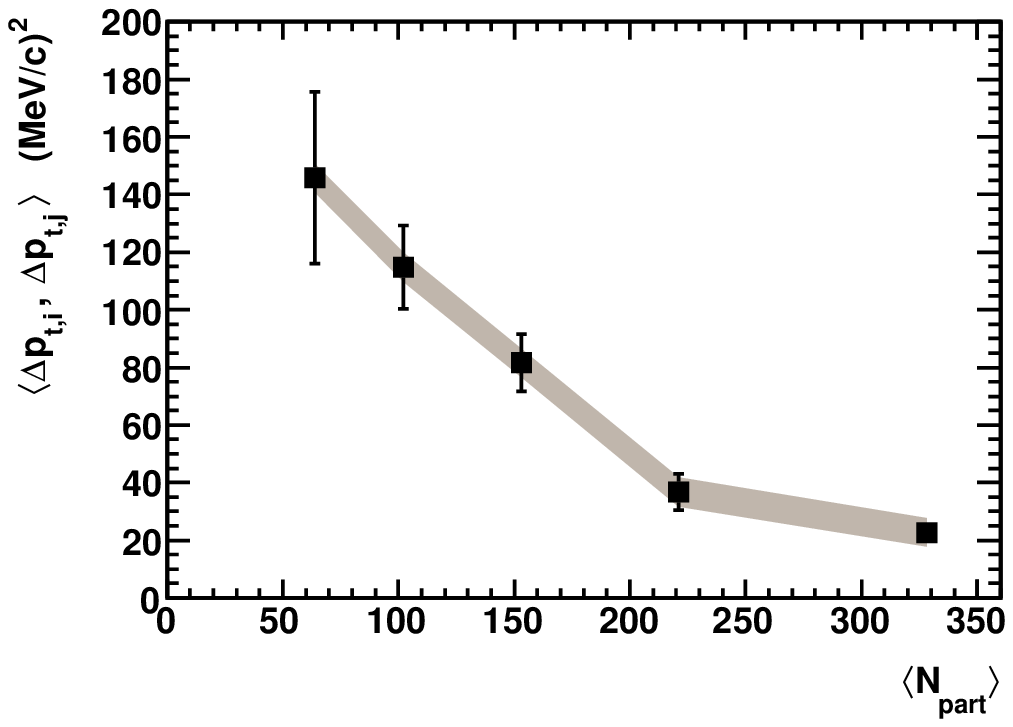}
\includegraphics[height=6cm]{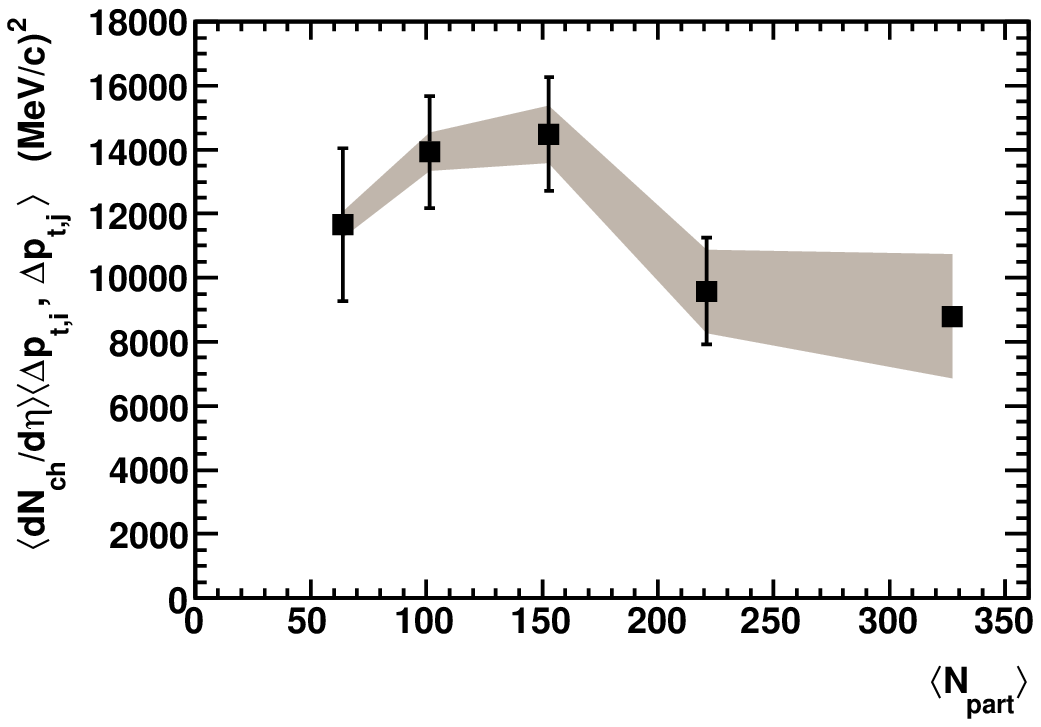}
\caption{The momentum correlator $\corr$ (left) and 
$\langle {\rm d}N_{\rm ch}/{\rm d}\eta \rangle \cdot \corr$ (right) 
from all particle combinations as function
of $\langle N_{\rm part}\rangle$. }
\label{fig-cent}       
\end{figure}

The centrality dependence of $\corr$ for all charged particles
is shown in Fig.\ref{fig-cent} and summarized in Table~\ref{tab:tab2}.
The correlator  $\corr$ exhibits a monotonic decrease, consistent
with recent results at SPS and 
RHIC~\cite{ceres-2,star-2}. This is partially explained 
by a dilution
effect which occurs in case of an independent superposition of colliding 
nucleons. To explore this behaviour in more detail we multiply the 
correlator by the charged particle multiplicity 
$\langle {\rm d}N_{\rm ch}/{\rm d}\eta \rangle$ in each bin, 
as shown in Fig.\ref{fig-cent}. Similar to previous observations
of related quantities at SPS and RHIC, 
$\langle {\rm d}N_{\rm ch}/{\rm d}\eta \rangle \cdot \corr$ 
is not independent of centrality
but rather indicates a non-monotonic shape with a maximum
at $\langle N_{\rm part}\rangle \approx 150$. 
Different mechanisms have been proposed as possible explanations
for this phenomenon, such as correlations due to jets and jet 
quenching~\cite{phenix-1}, onset of
thermalization~\cite{gavin}, elliptic flow~\cite{phenix-1,na49-1}, 
or enhanced fluctuations signalling
a geometric phase transition in the vicinity of the percolation 
point~\cite{ferr}. 
Also, a possible connection to enhanced multiplicity fluctuations observed
in the same centrality region~\cite{na49-mult} has been 
demonstrated~\cite{mrow-1}.
A comprehensive and convincing explanation, however, is not yet at hand. 

\subsection{Correlation analysis in $\Delta \eta - \Delta \phi$}
\begin{figure}
\centering
\includegraphics[height=5cm]{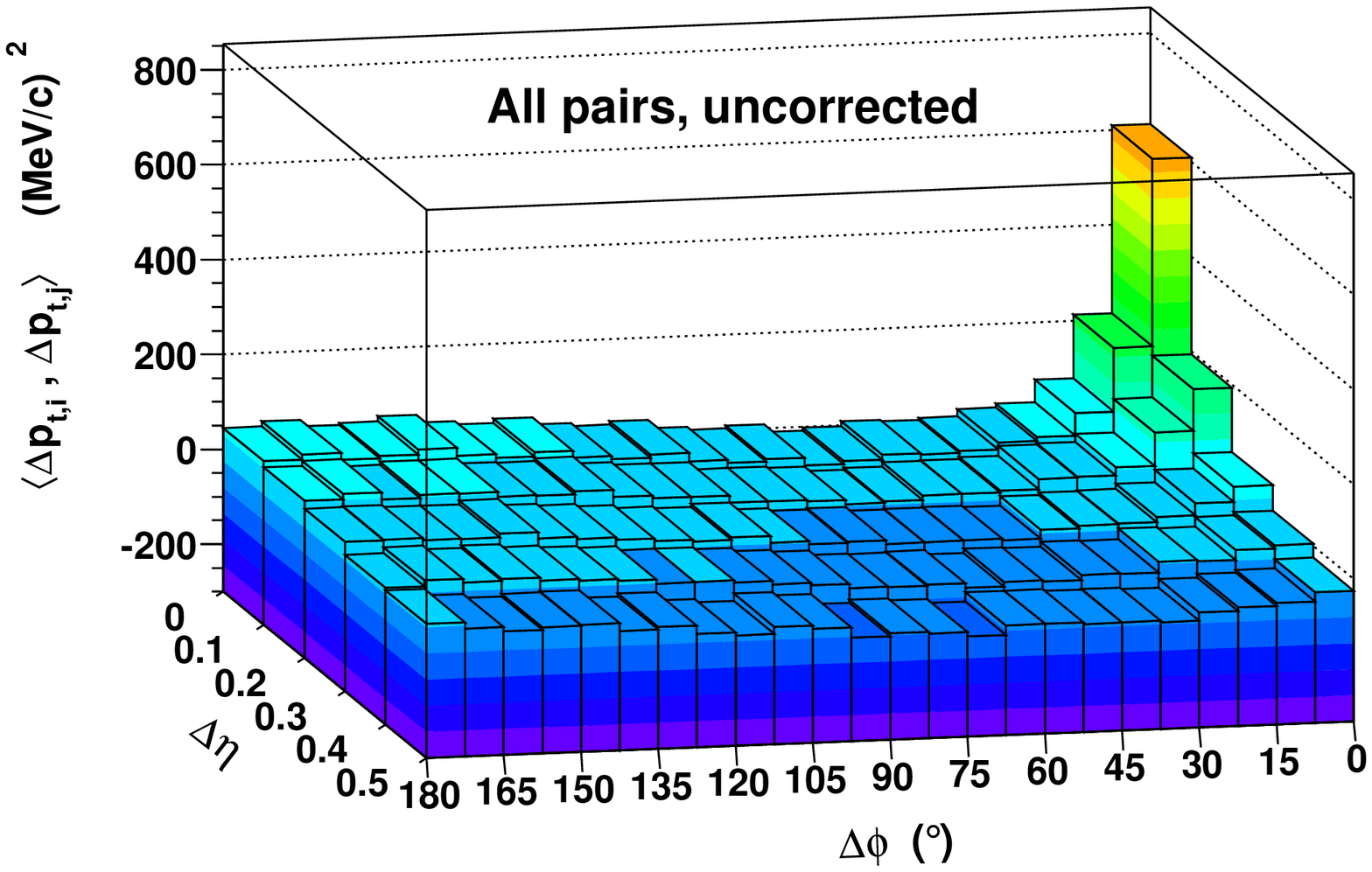}
\includegraphics[height=5cm]{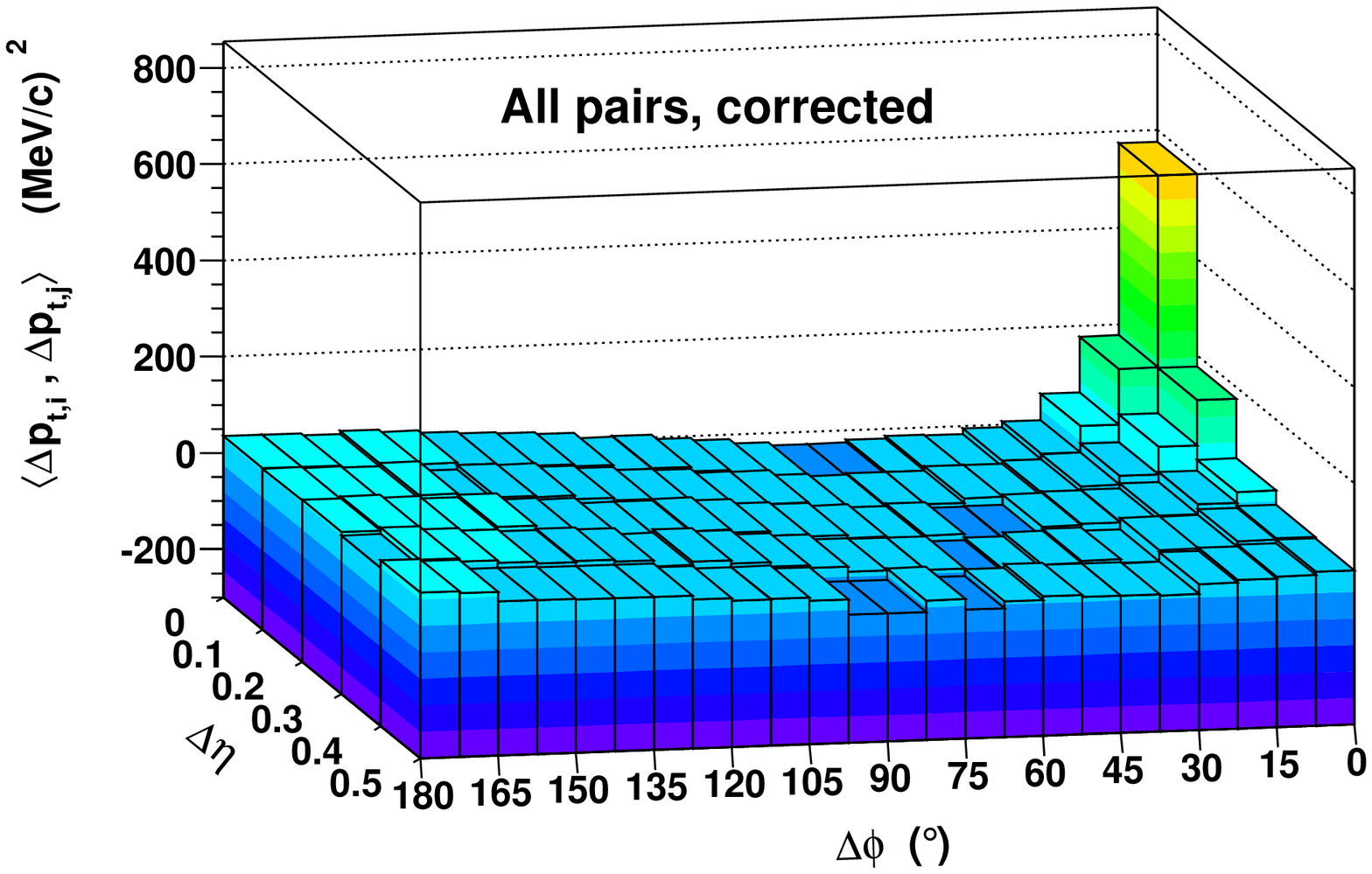}
\caption{The momentum correlator $\corr$ as function of $\Delta \eta$
and $\Delta \phi$. The left panel is before, the right panel after
subtraction of the mixed-event correlations. }
\label{map1-all}       
\end{figure}

To address this question in more detail, the scale dependence
of the momentum correlations in relative angular
space was investigated. This study will help to identify and
disentangle
different contributions to the observed fluctuation pattern.
Fig.\ref{map1-all} shows the momentum correlator $\corrdiff$
for all charged particle pairs in central Pb-Au collisions at 
158$A$ GeV/$c$. The left panel of Fig.\ref{map1-all} is uncorrected,
the right panel after subtraction of the mixed-event correlations.
The most pronounced feature is a strong positive correlation 
at small opening angle. The width of this peak is similar in
$\Delta \eta$ and $\Delta \phi$. At $\Delta \phi \approx \pi/2$,
a small but significant  negative correlation is observed,
turning into a positive plateau as $\Delta \phi$ approaches
180 degrees. The smooth decrease with $\Delta \eta$ observed 
in the uncorrected data is caused by acceptance effects and 
disappears after subtraction of the mixed-event data.

\begin{figure}
\centering
\includegraphics[height=5cm]{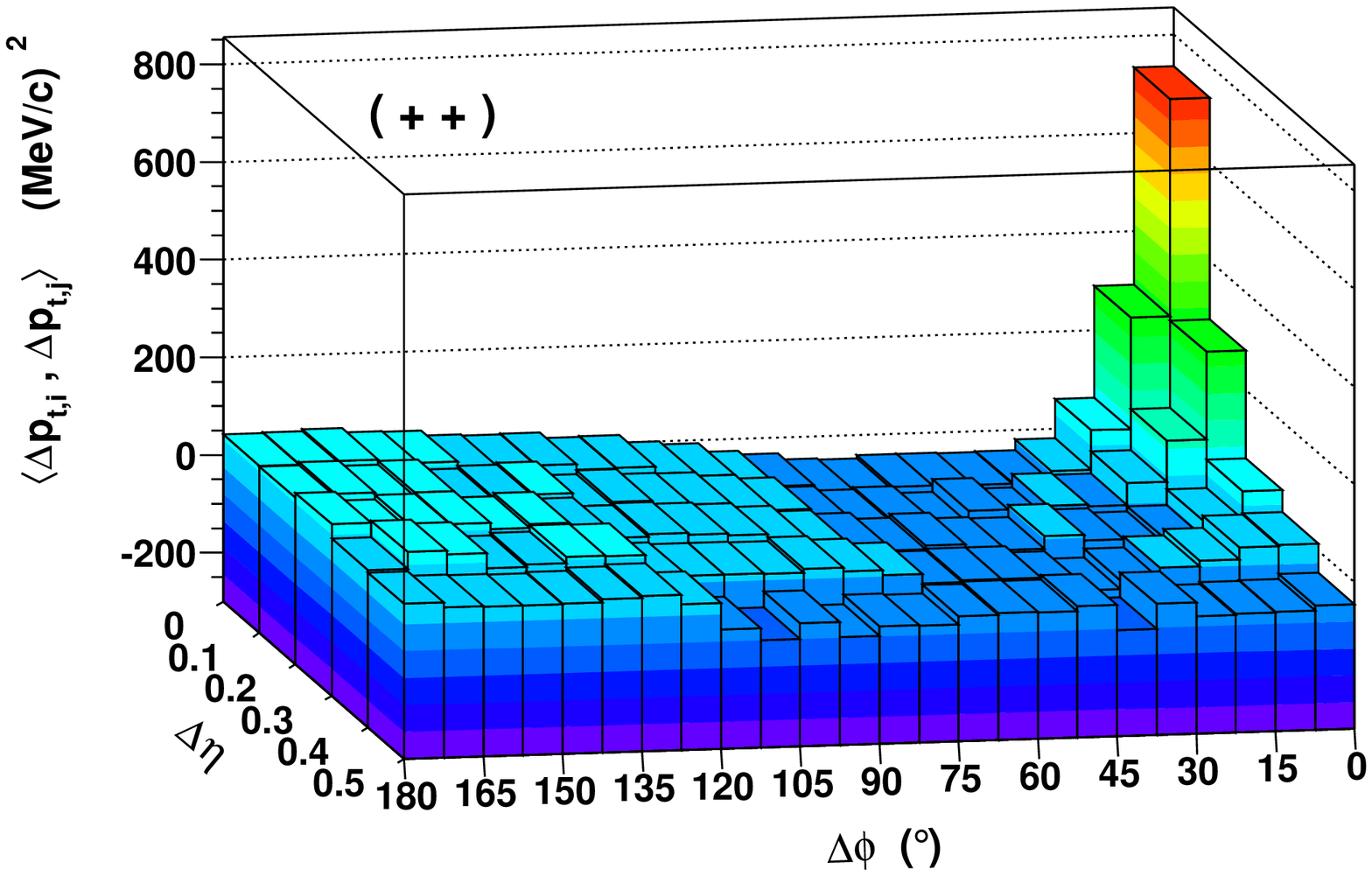}
\includegraphics[height=5cm]{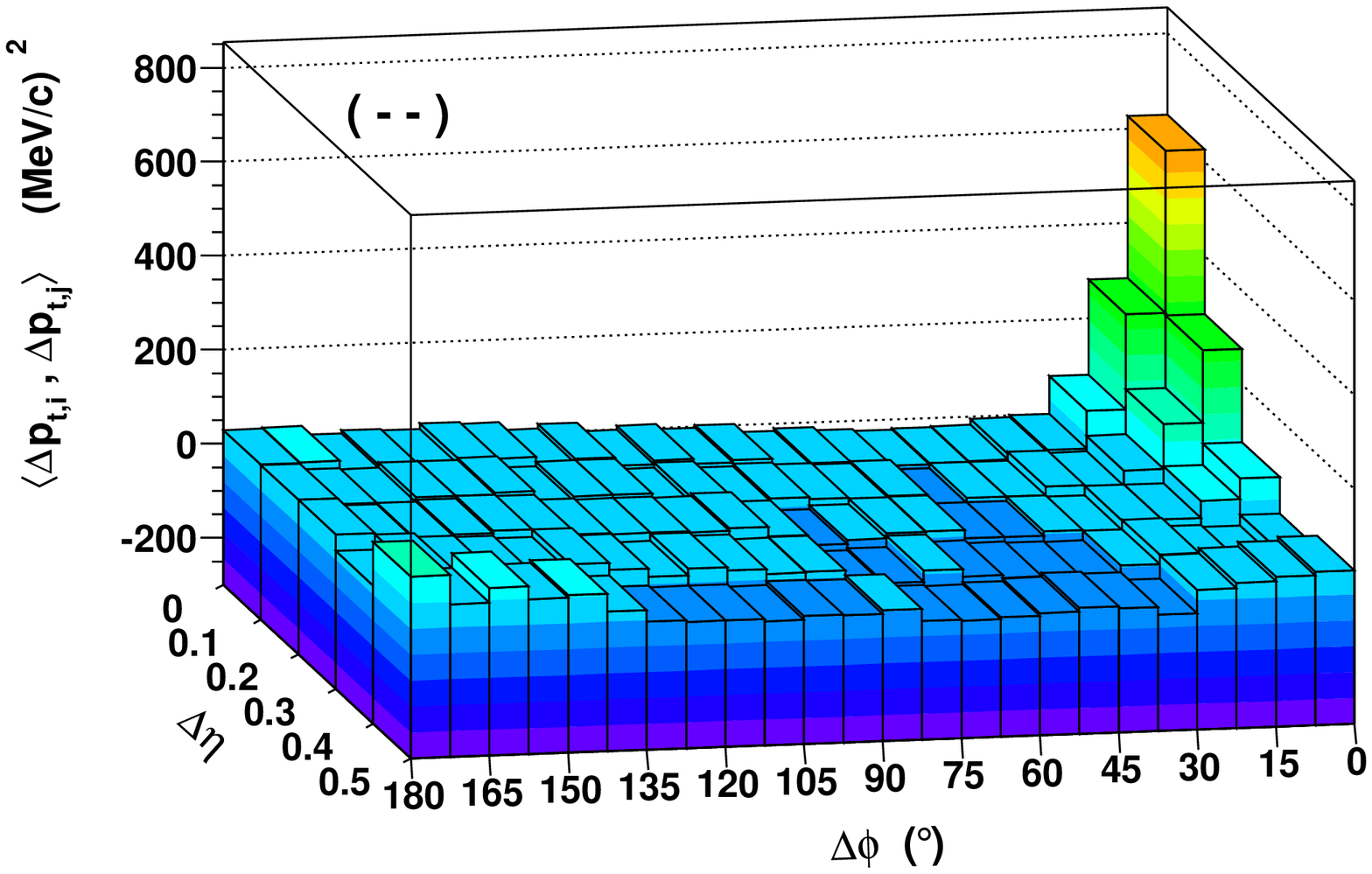}
\includegraphics[height=5cm]{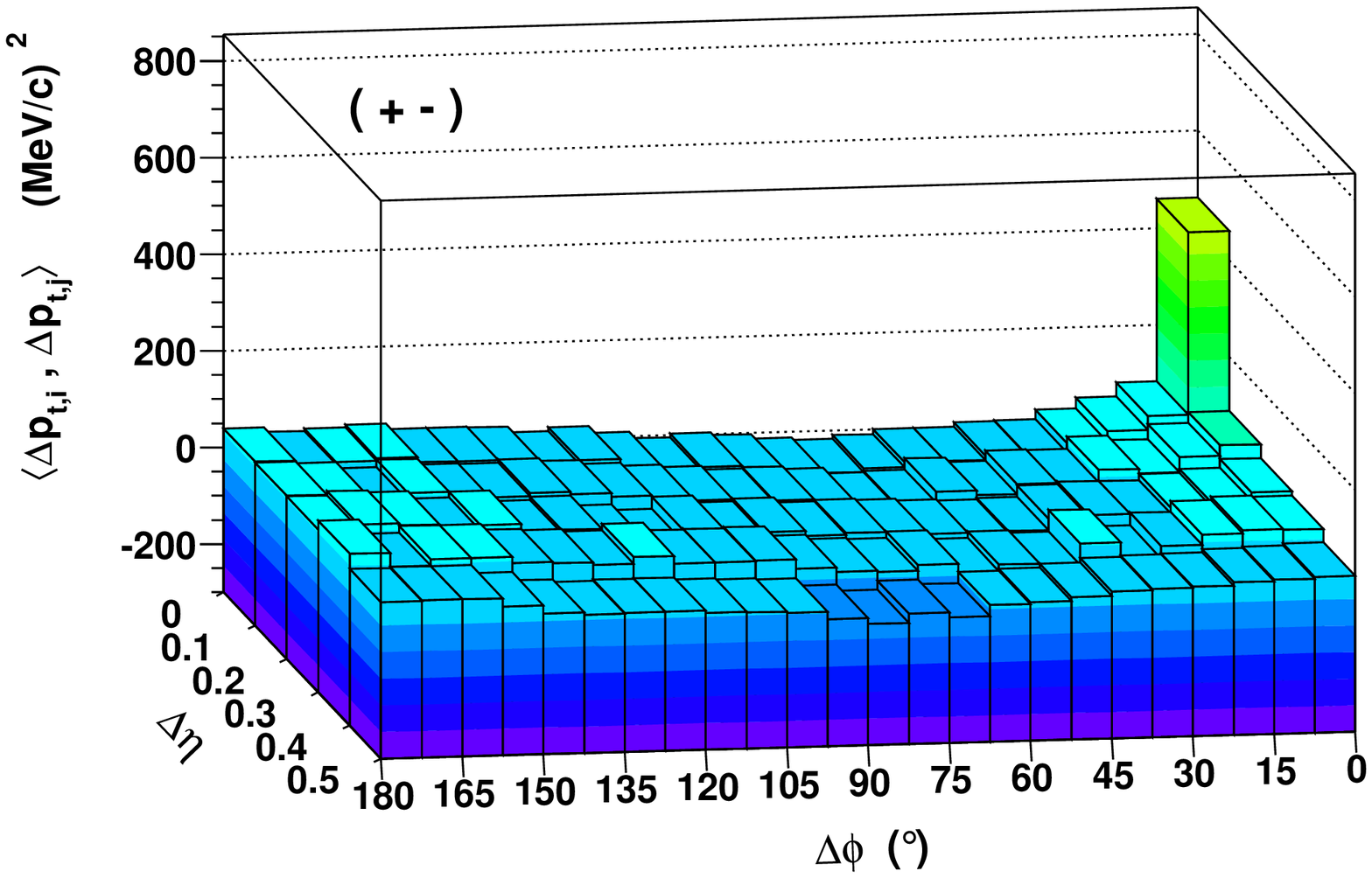}
\caption{The momentum correlator $\corr$ as function of $\Delta \eta$
and $\Delta \phi$. Upper left: positive pairs, upper right: negative pairs,
lower panel: unlike-sign pairs. }
\label{fig-charges}       
\end{figure}

The strong peak at small angles is most likely due to short range
correlations caused by quantum statistics and Coulomb final state
interactions. This hypothesis is substantiated when the momentum
correlations are calculated for different charge combinations.
In Fig.\ref{fig-charges}, the correlator 
$\corrdiff$ is shown for positive, negative
and unlike-sign pairs. The peak structure for positive and negative
pairs is very similar, pointing to HBT correlations as a likely
origin. 

In contrast, the peak is restricted to very small opening
angles for unlike-sign pairs, where no HBT correlations are
present. Here, the correlation is likely to be caused by attractive
Coulomb final state interactions. Also $e^+e^-$ pairs from conversions 
may contribute at very small opening angles.
It is interesting to note that the narrow spike is observed on top of 
a broad bump with much smaller amplitude, which has similar
width in $\Delta \eta$ and $\Delta \phi$.  
This may be caused by resonance decays, but also by correlations from
mini-jet fragmentation, where a stronger contribution to the
unlike-sign correlations as compared to the like-sign case is expected
as a consequence of charge ordering.

In the next step, we present the scale-dependence as function
of the centrality of the collision. For reasons of limited 
statistics in the minimum bias data set, we do not perform a
charge-dependent analysis of the centrality dependence. 
Furthermore, we neglect the $\Delta
\eta$ dependence and focus on $\Delta \phi$, integrating the
data over the $\Delta \eta$ acceptance.

\begin{figure}
\centering
\includegraphics[height=12cm]{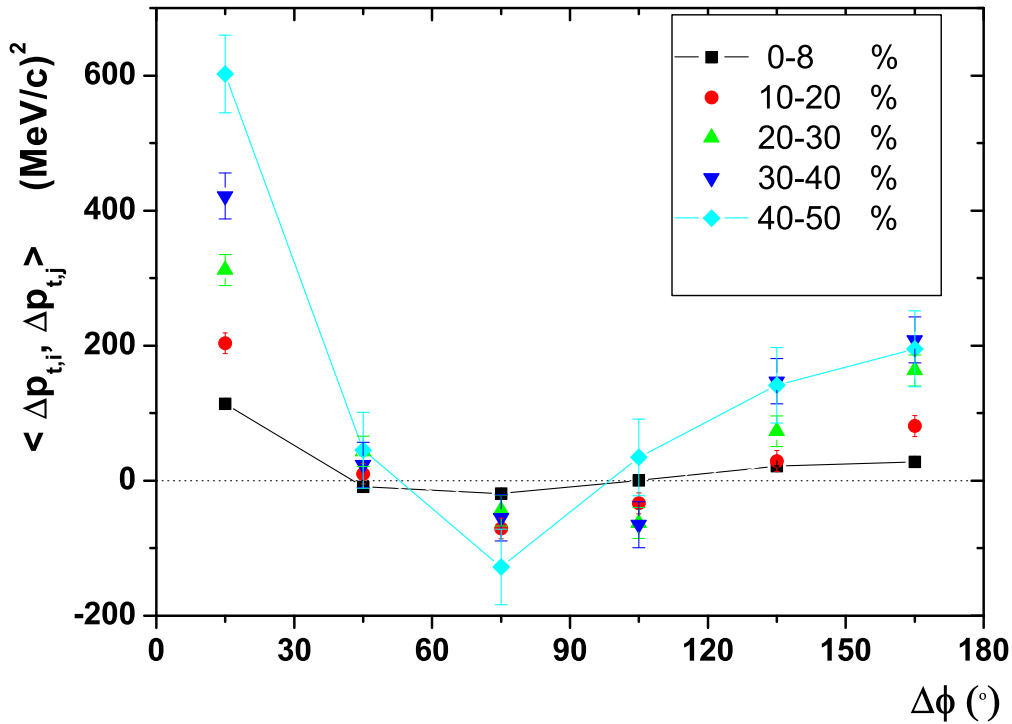}
\caption{The momentum correlator $\corr$ as function of 
 $\Delta \phi$ for different centrality classes. }
\label{cent-dphi}       
\end{figure}

The momentum correlator $\corr$ as function of $\Delta \phi$ for
different centralities is shown in Fig.\ref{cent-dphi}. The data
show a pronounced $\Delta \phi$ dependence, exhibiting a strong
peak in the first bin on top of a harmonic pattern with a 
second peak for back-to-back pairs. 
Significant anti-correlations are observed in the angular region
around $\Delta \phi \approx 1$.
The strength of the correlation decreases towards more central
events. 

\begin{figure}
\centering
\includegraphics[height=12cm]{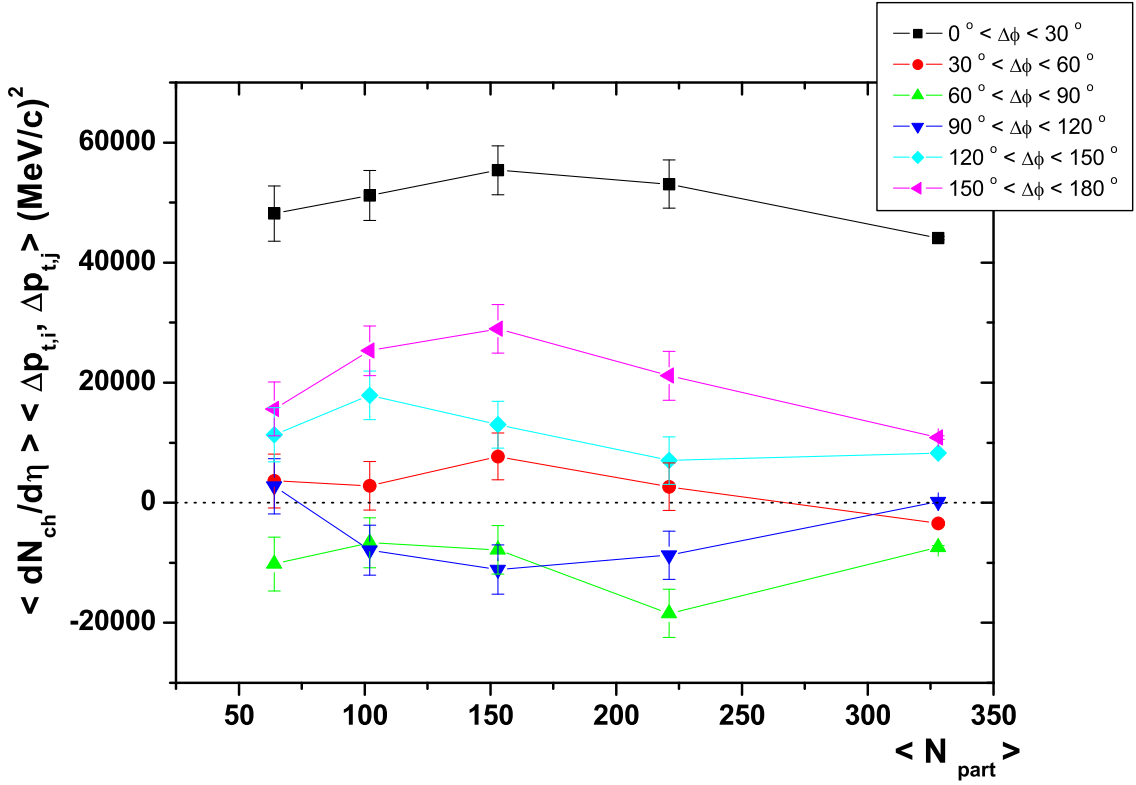}
\caption{The momentum correlator $\langle {\rm d}N_{\rm ch}/{\rm d}\eta \rangle \cdot
\corr$ as function of $\langle N_{\rm part}\rangle$ for 
different $\Delta \phi$
intervals. }
\label{dphi-cent}       
\end{figure}

Fig.\ref{dphi-cent} shows 
$\langle {\rm d}N_{\rm ch}/{\rm d}\eta \rangle \cdot \corr$ as function
of $\langle N_{\rm part}\rangle$ in different regions of $\Delta \phi$.  
The non-monotic behaviour observed in the scale-independent analysis
turns out to exhibit a significant dependence on $\Delta \phi$. As
expected, the correlation is largest at the smallest $\Delta \phi$.
In this angular range, the data show a maximum around 
$\langle N_{\rm part}\rangle \approx 150$, however, 
the data seem to be biased by a large offset possibly
arising from a centrality-independent 
contribution from HBT and Coulomb effects.
With increasing $\Delta \phi$, the correlation strength decreases
and turns negative around $\Delta \phi = \pi/2$. In this angular
range, the data indicate a {\em minimum} in semi-central events.
As $\Delta \phi$ increases
further, the correlations become again positive with a maximum
around $\langle N_{\rm part}\rangle \approx 150$.
Note that in the angular range $30^{\circ} < \Delta \phi < 60^{\circ}$
the result for the correlator is consistent with zero at all 
centralities.

The angular pattern of the momentum correlations
as well as the extrema observed in semi-central events suggest
a connection to elliptic flow as a possible origin.
In the following, an estimate of the expected flow
contribution to the momentum correlations, based on measurements
of the elliptic flow strength $v_2$ is presented.
The Fourier coefficient $v_2$ of the second order harmonic 
of the azimuthal anisotropy of particle emission has been
measured in Pb-Au events at 158$A$ GeV/$c$ as function of
$p_t$~\cite{jovan-qm05}. 
We employ a parametrization of the $p_t$ dependence
of $v_2$ in the most central event class (0-8\%).
Two alternative approaches have been tested to 
estimate its contribution to $\corr$. 
In the first method, a weight $f_{ij}$ is introduced for 
each particle pair $i,j$ which depends on the parametrized
$v_2$ value at the particle's $p_t$:
\begin{equation}
f_{ij} = 1+2v_2(p_{t,i})v_2(p_{t,j})\cos(2|\phi_i - \phi_j|).
\end{equation}
Employing the mixed-event technique, the correlator $\corr_{\rm mixed, flow}$
has been evaluated by weighting each pair of particles
from different events with $f_{ij}$ and normalization by the
sum over all weights $F_{ij}=\sum{f_{ij}}$: 
\begin{equation}
\corr_{\rm mixed, flow}=
\frac{1}
{F_{ij}}
\sum_{i,j}f_{ij}(p_{ti}-\overline{p_t})
(p_{tj}-\overline{p_t}) ,
\end{equation}
with particles $i$ and $j$ from different events.
To account for the single particle acceptance, the unweighted 
mixed-event correlator has been subtracted 
\begin{equation}
\corr_{\rm flow}= \corr_{\rm mixed, flow} - \corr_{\rm mixed}.
\end{equation}

\begin{figure}
\centering
\includegraphics[height=9cm]{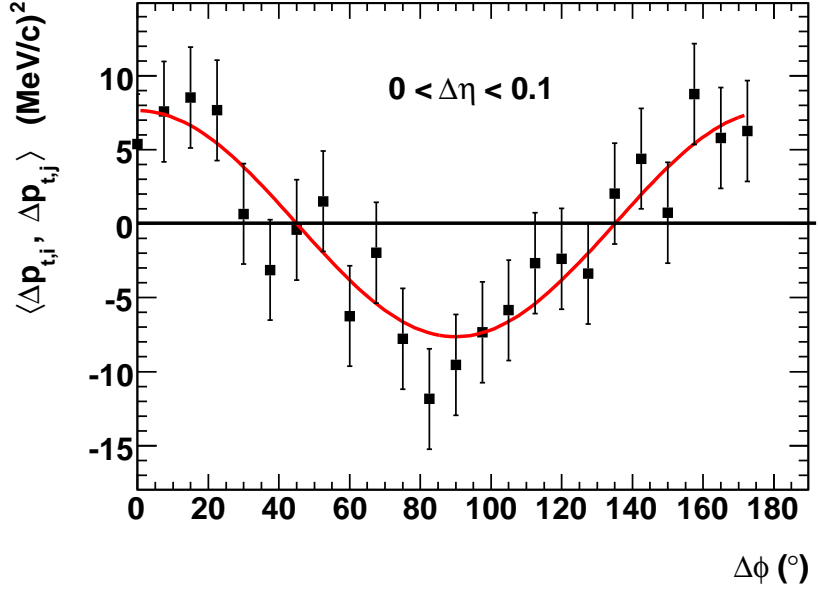}
\caption{Estimated contribution of elliptic flow to the momentum 
correlator $\corr$ as function of $\Delta \phi$, obtained from 
a Monte Carlo simulation.
The results are parametrized by a harmonic function.}
\label{corr-flow}       
\end{figure}

\begin{figure}
\centering
\includegraphics[height=10cm]{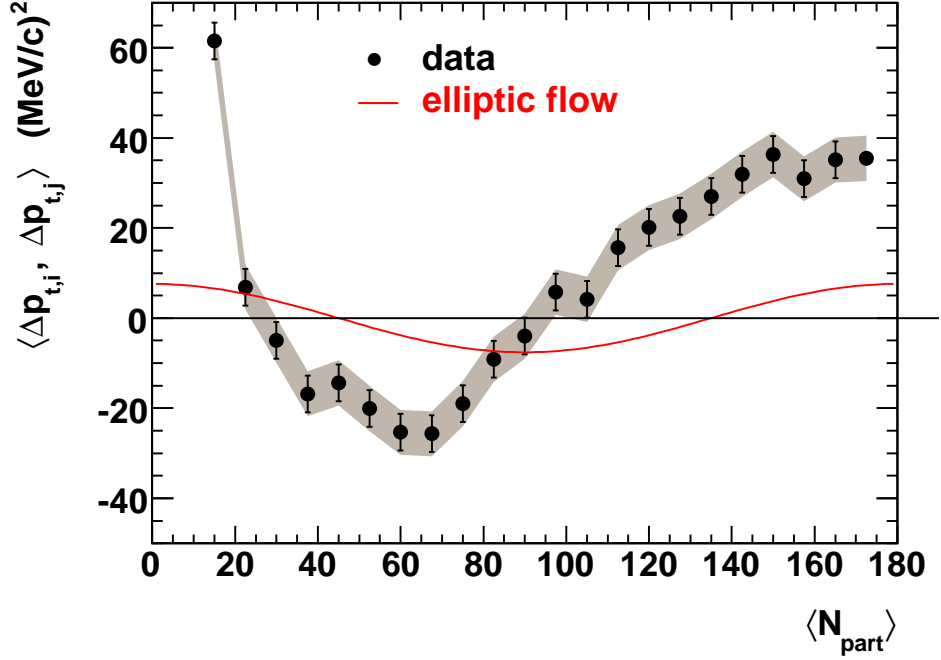}
\caption{The momentum correlator $\corr$ as function of 
$\langle N_{\rm part}\rangle$
in $0 < \Delta \eta < 0.1$ compared to the expectation from elliptic
flow (curve).}
\label{flow-corrdata}       
\end{figure}

The resulting correlation as function of $\Delta\phi$ in the 
range $0<\Delta\eta<0.1$ is shown in Fig.\ref{corr-flow}. As 
expected, the flow contribution has a characteristic $\cos(2\Delta\phi)$-
shape. However, the estimated flow contribution fails to describe the
measured correlation in shape and in magnitude, as demonstrated
in Fig.\ref{flow-corrdata}. The same quantitative result has been 
obtained by a Monte-Carlo method, where parametrizations of the 
measured $p_t$ distributions and $v_2$ values have been used to calculate
the correlator $\corr_{\rm flow}$. 

\subsection{Two-particle $p_t$ correlations}
In this section, we try to obtain further insight into the origin of
the observed $p_t$ correlations by the investigation of their $p_t$ dependence.
Non-zero $p_t$ correlations as observed in the measure $\corr$ must be related 
to (anti-)correlated particle production at the corresponding angular scale. 
A localization of correlations in transverse momentum space 
may help to distinguish between different mechanisms which could 
possibly lead to such correlations.
To this end, we study two-particle correlations in the $(x(p_t)_1, x(p_t)_2)$-
plane, where the cumulative $p_t$ variable $x(p_t)$ is defined 
as~\cite{na49-1,bial-gadz}:
\begin{equation}
x(p_t) = \frac{\int_{0.1~{\rm GeV}/c}^{p_t}\rho(p_t'){\rm d}p_t'}
{\int_{0.1~{\rm GeV}/c}^{1.5~{\rm GeV}/c}\rho(p_t'){\rm d}p_t'}.
\end{equation}

\begin{figure}
\centering
\includegraphics[height=7cm]{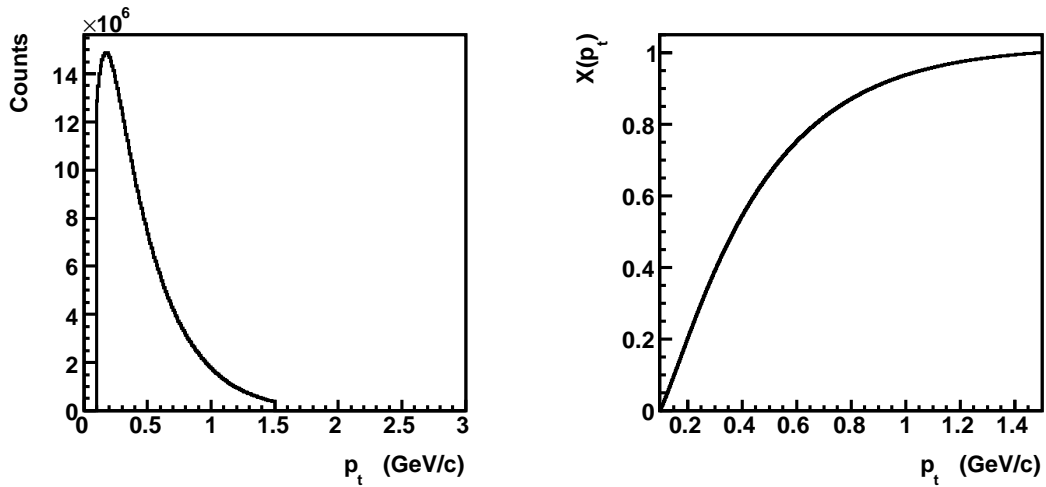}
\caption{The inclusive $p_t$ distribution (left) and the cumulative $p_t$ 
variable $x(p_t)$ (right).}
\label{ptinc-xpt}       
\end{figure}

In Fig.\ref{ptinc-xpt}, $x(p_t)$ is shown as function of $p_t$ 
next to the inclusive $p_t$ distribution $\rho(p_t)$.
Due to the limited statistics of our peripheral data set we focus on
central collisions (0-5\%) in the following discussion.

For the study of two-particle $p_t$ correlations, the $x(p_t)$-values
of particle pairs $(x(p_t)_1,x(p_t)_2)$ are filled into 
two-dimensional arrays. The same procedure has been applied to mixed-event 
pairs to study residual $p_t$ correlations caused by detector acceptance 
effects. Such correlations may occur in differential analyses, if local
deviations from the inclusive $p_t$ distributions are present.
For the final results presented below, the real-event distributions 
have been corrected for detector effects through division by the 
corresponding mixed-event distributions.

\begin{figure}
\centering
\includegraphics[height=12cm]{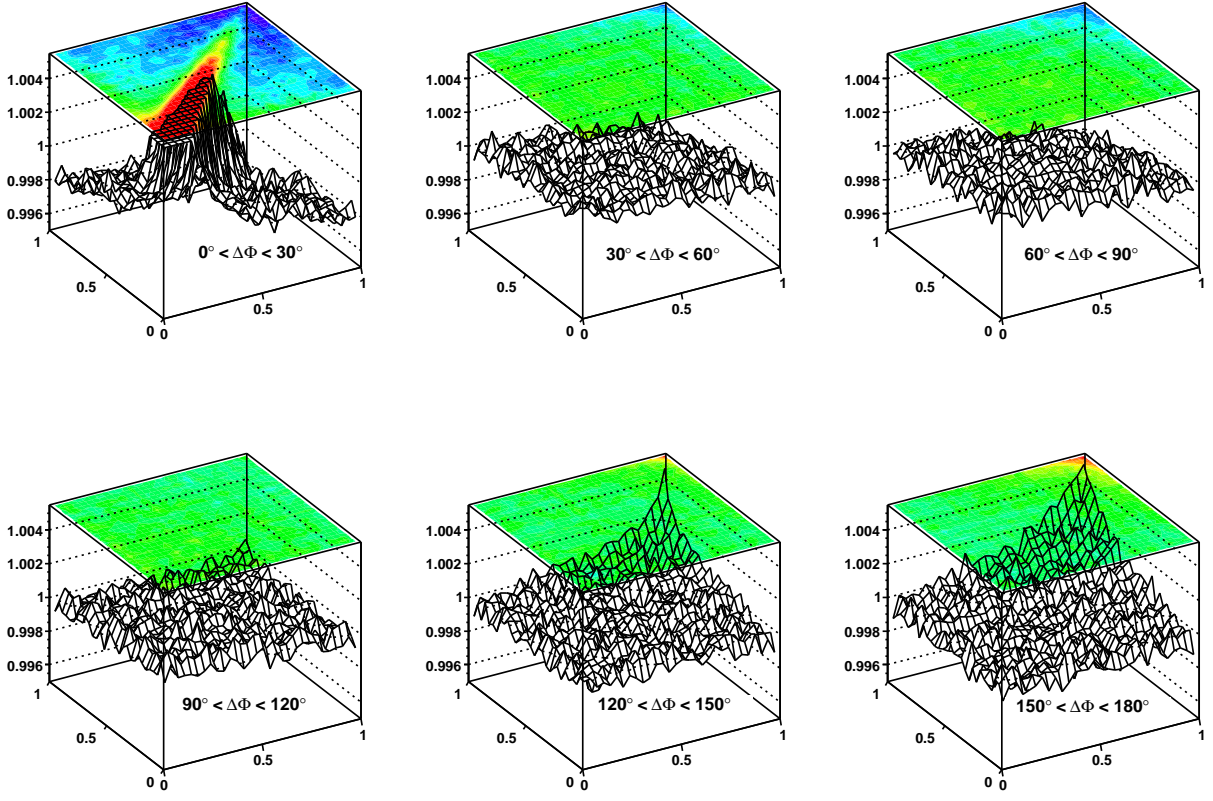}
\caption{Two-particle correlations as function of $(x(p_t)_1,x(p_t)_2)$ in 
different regions of $\Delta \phi$. No charge selection has been applied.}
\label{xpt1}       
\end{figure}

\begin{figure}
\centering
\includegraphics[height=10cm]{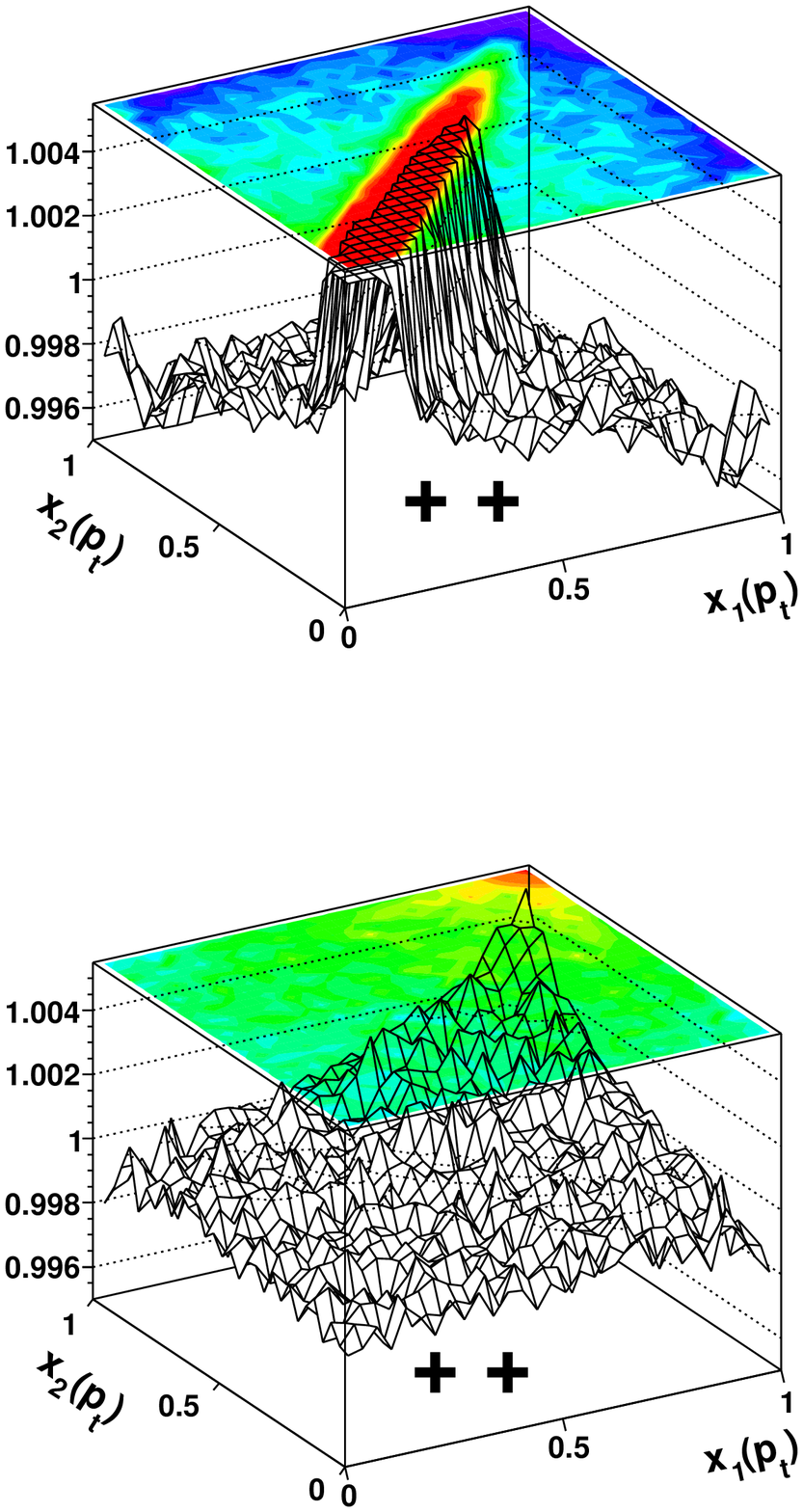}
\includegraphics[height=10cm]{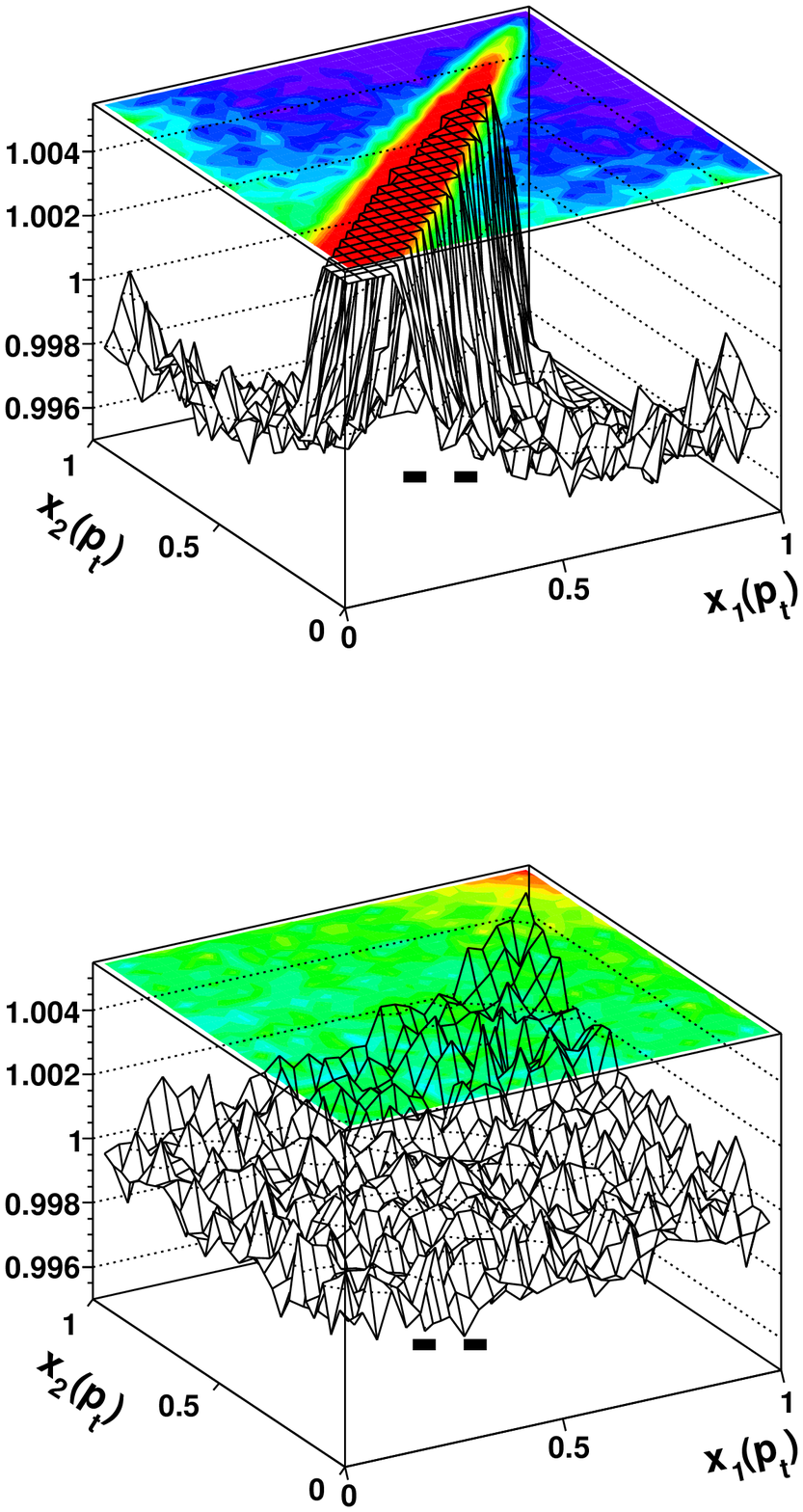}
\includegraphics[height=10cm]{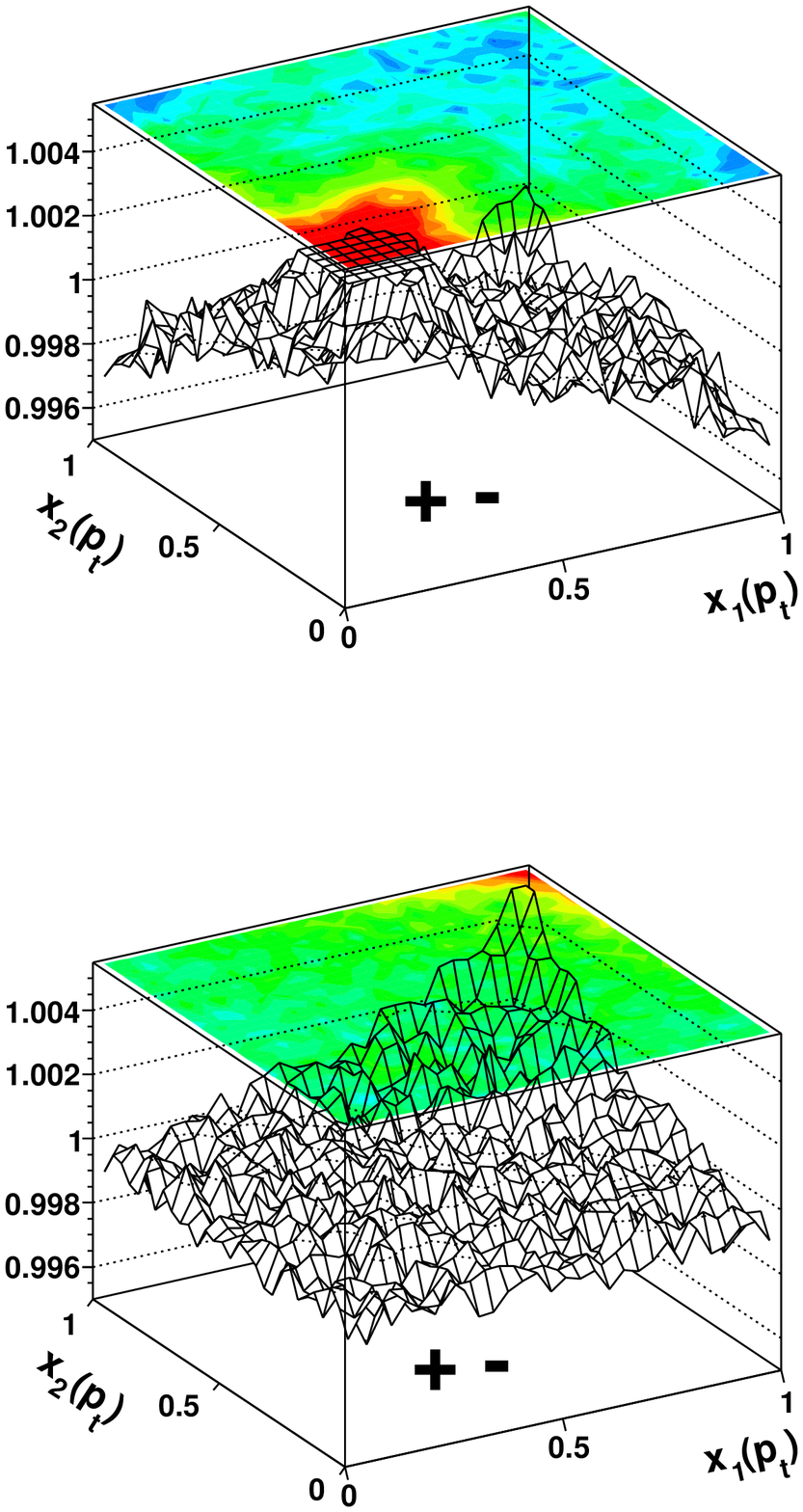}
\caption{Two-particle correlations as function of $(x(p_t)_1,x(p_t)_2)$ in 
$0^{\circ}<\Delta \phi<30^{\circ}$ (upper row) and 
$150^{\circ}<\Delta \phi<180^{\circ}$ (lower row). 
Results for positive, negative and
unlike-sign particle pairs are shown separately.}
\label{xpt2}       
\end{figure}

In Fig.\ref{xpt1}, the two-particle correlation function from central
Pb-Au events (0-5\%) is shown
in the $(x(p_t)_1,x(p_t)_2)$-plane for different values of the relative 
azimuthal separation $\Delta\phi$ of the pair. For small $\Delta \phi$ 
($0< \Delta \phi < 30^{\circ}$) we observe a strong positive correlation
along the $x_1 \approx x_2$ - diagonal, most pronounced in the low 
$p_t$-region. This is consistent
with short-range correlations of particles with small momentum differences as
induced by quantum statistics and Coulomb final state interactions. 
The short-range contribution vanishes for large angular separations. 
At $\Delta \phi \approx 90^{\circ}$ an anti-correlation at large transverse
momenta appears, which turns to a substantial positive correlation as 
the angular separation approaches $\Delta \phi = 180^{\circ}$.

The charge-dependent correlation functions in the $(x(p_t)_1,x(p_t)_2)$-plane
for small ($0< \Delta \phi < 30^{\circ}$) and large ($150< \Delta \phi < 
180^{\circ}$) azimuthal separations of the particle pair are shown in 
Fig.\ref{xpt2}. The correlations of positive and negative particle
pairs look similar at small $\Delta \phi$, pointing to a common mechanism
such as HBT. On the other hand, the correlation of unlike-sign particles is 
restricted to significantly lower $p_t$, as expected from Coulomb interaction
and contaminations from $e^+e^-$ - pairs. The positive high-$p_t$ correlation
at large $\Delta \phi$ appears very similar for all charge combinations.

From this pattern, the following picture arises: the strong positive 
values of $\corr$ at small $\Delta \phi$ are dominated by HBT and Coulomb
correlations. In contrast, the characteristic structure in Fig.\ref{cent-dphi} 
at $\Delta \phi > 30^{\circ}$ is caused by (anti-) correlations at high $p_t$
which can not be attributed to elliptic flow. 
Using the flow model described before, we found that the value 
of $v_2$ needed to be three times larger than the experimental $v_2$ value
in order to account for the observed correlations.
Similar conclusions have been
drawn from an analysis of azimuthal correlations with respect to a 
high-momentum trigger
particle at the same beam energy, 
where the observed correlation pattern is significantly larger
than the contribution expected from elliptic flow~\cite{mat-qm05}. 
Such triggered correlation 
functions are usually attributed to jet correlations arising from the 
fragmentation of hard-scattered partons with large transverse momentum.
Related studies of two-particle $p_t$-correlations have been performed by
STAR at RHIC~\cite{tt3}, where similar observations have been discussed in the
context of initial state semi-hard parton scattering and minijet dissipation
in tge medium.

\subsection{Summary}

Based on the large statistics data set of Pb-Au collisions at 158$A$ GeV/$c$,
recorded with the CERES Time Projection Chamber at the CERN-SPS, a novel 
analysis of transverse momentum correlations has been performed. 
In a differential analysis of two-particle correlations as function of
the charge sign and
the relative angular separation of the particle pairs, the scale dependence
of $p_t$ correlations was determined. We found that the previously observed
non-statistical transverse momentum fluctuations can be decomposed into
different contributions.
First, a short range component at small angles and low transverse momentum
has been identified,
which is most likely due to HBT (in the case of like-sign particle pairs) and
Coulomb interactions and conversions (in the case of unlike-sign particle
pairs). Second, a pronounced long-range component in $\Delta \phi$ is observed, 
located at large transverse momentum and exhibiting
minima and maxima, which are distinctly different from those expected for elliptic flow. 
A possible interpretation is that this component originates in kinematic correlations arising
from jet fragmentation of energetic partons. 

Beyond these contributions, we could not identify additional sources 
of particle correlations. In particular, no trace of non-trivial event-by-event
fluctuations has been found but it may well be overwhelmed by the dominant 
contributions discussed in this paper. We note, however, that the angular
range $30^{\circ} < \Delta \phi < 60^{\circ}$ is essentially free of
such contributions. In this region, the observed correlations are
consistent with zero.
Based on the studies presented here, we suggest 
to optimize the sensitivity of future investigations to this angular
range as it appears to offer the cleanest window for the observation
of non-trivial fluctuations connected to the critical point. 

\begin{table}
\caption{\label{tab:tab1}
Results of a scale-independent analysis of central Pb-Au events at
158$A$ GeV/$c$ ($\sigma/\sigma_{\rm geo} = 0-8\%$).}
\begin{tabular}{ccccc}
~& Positive pairs & Negative pairs & All pairs & Unlike-sign pairs \\
\hline
$\corr$ (MeV$^2$/$c^2$) & 
	$21.59\pm0.63$ & $26.63\pm0.61$ & $22.71\pm0.32$ & $24.71\pm0.43$  \\
$\sigma_{pt,{\rm dyn}}^2$ (MeV$^2$/$c^2$) & 
	$20.65\pm0.61$ & $26.16\pm0.54$ & $21.98\pm0.44$ & ~ \\
$\Sigma_{p_t}$ (\%) & 
	$0.95\pm0.01$ & $1.24\pm0.01$ & $1.04\pm0.01$ & ~ \\
$n$ & 
	9009425 & 9009425 & 10003672 & 9009425 \\
$\langle N \rangle$ & 
	$84.21\pm0.01$ & $70.63\pm0.01$ & $154.83\pm0.01$ & ~ \\
$\overline{p_t}$ (MeV/$c$) & 
	$479.78\pm0.01$ & $414.10\pm0.01$ & $449.82\pm0.01$ & ~ \\
\end{tabular}
\end{table}

\begin{table}
\caption{\label{tab:tab2}
Centrality dependence of mean $p_t$ correlation results in Pb-Au at 158$A$ GeV/$c$.
No charge selection was applied.}
\begin{tabular}{ccccccc}
$\sigma/\sigma_{\rm geo}$ & $\langle N_{\rm part} \rangle$ & $n$ & $\langle N \rangle$ & 
$\overline{p_t} $ & 
$\corr$ & $\sigma_{pt,{\rm dyn}}^2$ \\
 & & & & (MeV/$c$) & (MeV$^2$/$c^2$) & (MeV$^2$/$c^2$)\\
\hline
40-50\% & 64  & 38197 & $32.72\pm0.06$  & $436.17\pm0.25$ & $145.79\pm29.73$ & $129.62\pm35.69$  \\
30-40\% & 102 & 39763 & $50.07\pm0.07$  & $441.57\pm0.2 $ & $114.81\pm14.41$ & $95.81\pm18.81$\\
20-30\% & 153 & 39496 & $73.37\pm0.08$  & $445.31\pm0.17$ & $81.61\pm9.99$ & $68.62\pm12.7$ \\
10-20\% & 221 & 40146 & $104.81\pm0.1$  & $448.52\pm0.14$ & $36.74\pm6.39$ & $35.52\pm8.61$  \\
0-8\%   & 328 & 10003672 & $154.83\pm0.01$ & $449.82\pm0.01$ & $22.71\pm0.32$ & $21.98\pm0.44$ \\
\end{tabular}
\end{table}

\section{Acknowledgements} \nonumber

This work was supported by GSI, the German BMBF, the Virtual Institute VI-SIM
of the German Helmholtz Association, the Israel Science Foundation,
the Minerva Foundation, and by the Grant Agency and Ministry of Education
of the Czech Republic.


\end{document}